\documentclass[11pt]{cernrep}

\usepackage{graphicx}
\usepackage{here}

\newcommand{\quarter}{\frac{1}{4}}

\newcommand{\half}{\frac{1}{2}}

\newcommand{\ds}{$D^*_{sJ}$(2317)$^+$}

\begin{document}

\title{MESON SPECTRA FROM AN EFFECTIVE LIGHT CONE QCD-INSPIRED MODEL}

\author{Shan-Gui Zhou}
\institute{Max-Planck-Institut f\"ur Kernphysik, 69029 Heidelberg, Germany\\
           School of Physics, Peking University, Beijing 100871, China\\
           Institute of Theoretical Physics, Chinese Academy of Sciences,
           Beijing 100080, China \\
           Center of Theoretical Nuclear Physics, National Lab. of
           Heavy Ion Accelerator, Lanzhou 730000, China}

\maketitle

\begin{abstract}
I present some recent applications of a light cone QCD-inspired
model with the mass squared operator consisting of a harmonic
oscillator potential as confinement in the meson spectra. The
model gives an universal and satisfactory description of both
singlet and triplet states of $S$-wave mesons. In the present work
$P$-wave $D_s$ mesons are also investigated. The mass of the
recently found meson, \ds\ is reproduced fairly well by this
simple model.
\end{abstract}

\section{INTRODUCTION}
\label{sec:intro}

In the effective light cone QCD theory~\cite{BRO98,PAU00} the
lowest Fock component of the hadron wave function is an
eigenfunction of an effective mass squared operator with
constituent quark degrees of freedom and parameterized in terms of
an interaction which contains a Coulomb-like potential and a
Dirac-delta term. All higher Fock-state components of the hadron
light-front wave function can be constructed recursively from the
lowest one. The interaction in the mass operator comes from an
effective one-gluon-exchange where the Dirac-delta term
corresponds to the hyperfine interaction.

The masses of the ground state of the pseudoscalar mesons and in
particular the pion structure~\cite{FRE01} were described
reasonably, with a small number of free parameters, which is only
the canonical number plus one---the renormalized strength of the
Dirac-delta interaction. This model was also extended to include
the confining interaction and used to study the $S$-wave meson
spectra universally~\cite{FRE02,FRE02b}. In the light cone
framework (mass squared operator appears in the Hamiltonian), the
confining harmonic oscillator potential gives a natural
explanation of the observation of an almost linear relationship
between the mass squared of excited states with radial quantum
number $n$~\cite{ANI00}. We note that quite recently, another
harmonic oscillator approach was also used in the description of
scalar mesons successfully~\cite{KLE03}.

Recently a new narrow resonance \ds\ was observed by the BABAR
Collaboration~\cite{BABAR03} and confirmed by the CLEO
Collaboration~\cite{CLEO03}. Both the small mass and the small
width are in conflict with predictions of many
models~\cite{GOD85,GK91,DE01}. Since then much discussion has been
made concerning this new meson state. Some authors suggest that it
is a four quark meson~\cite{TER03}. Many others argue that it is
still within the normal picture of meson, namely, consisting of
two constituent quarks~\cite{BEV03}. The effective light cone
Hamiltonian model---the light cone harmonic oscillator (LCH) model
proposed in~\cite{FRE02,FRE02b} was also applied to this new
meson~\cite{ZHOU03}. The mass of \ds\ could be reproduced quite
accurately by the LCH model without changing the parameters fixed
previously.

The details of the application of this model in $S$-wave mesons
were published in \cite{FRE02,FRE02b}. In the present paper, I
will concentrate on $P$-wave $D_s$ mesons. Compared
to~\cite{ZHOU03}, a more reasonable spin-orbit interaction was
adopted here. In section~\ref{sec:swave}, I briefly review the
formalism of this model. The discussion on \ds\ is presented in
section~\ref{sec:ds}. Finally I will give a summary.

\section{THE LIGHT CONE HARMONIC OSCILLATOR MODEL}
\label{sec:swave}

In the effective light cone QCD theory the bare mass operator
equation for the lowest Fock-state component of a bound system
consisting of a constituent quark and antiquark with masses $m_1$
and $m_2$, is described as~\cite{BRO98,PAU00}
\begin{eqnarray}
 M^2 \psi (x,{\vec k_\perp})
 & = &
  \left[ \frac{{\vec k_\perp}^2+m^2_1}{x}
       + \frac{{\vec k_\perp}^2+m^2_2}{1-x}
  \right]
  \psi (x,{\vec k_\perp})
\nonumber \\
 &   & \mbox{}
  - \int \frac{dx' d{\vec k'_\perp}\theta(x')\theta (1-x')}
              {\sqrt{x(1-x)x'(1-x')}}
  \left( \frac{4m_1m_2}{3\pi^2} \frac{\alpha}{Q^2}
       - \lambda - W_{\rm{conf}}(Q^2)
  \right)
  \psi (x',{\vec k'_\perp}) \ ,
\label{p1}
\end{eqnarray}
where $M$ is the mass of the bound-state and $\psi$ is the
projection of the light-front wave-function in the quark-antiquark
Fock-state. The confining interaction is included in the model by
$W_{\rm{conf}}(Q^2)$. The momentum transfer $Q$ is the space-part
of the four momentum transfer and the strength of the Coulomb-like
potential is $\alpha$. The singular interaction is active only in
the pseudoscalar meson channel with $\lambda$ as the bare coupling
constant.

The mass operator equation could be transformed to the instant
form representation~\cite{PAU00b} and further simplified by
omitting the Coulomb term to the form~\cite{FRE02,FRE02b}
\begin{eqnarray}
 \left( M^2_{\rm ho} + g \delta (\vec r) \right) \varphi (\vec r)
 = M^2 \varphi (\vec r)
 \ ,
 \label{mass3}
\end{eqnarray}
where the bare strength of the Dirac-delta interaction is $g$, and
the mass squared operator is
\begin{equation}
 M^2_{\rm ho}
 = 2 m_t
   \left( -\frac{\vec{\nabla}^2}{2m_r}
          +\half m_t - c_0 + \half c_2 r^2
          +\kappa \vec L \cdot
            \left[\frac{\vec\sigma_1}{m_1^2}+\frac{\vec\sigma_2}{m_2^2}\right]
        \right)
 \ ,
 \label{mass4}
\end{equation}
where $m_t = m_1+m_2$ and $m_r = m_1m_2/(m_1+m_2)$. The harmonic
oscillator potential is introduced as a confinement with $c_0$ and
$c_2$ being two universal parameters valid for all mesons. The
phenomenological spin-orbit term~\cite{GOD85} is included for
$L\ne 0$ states which is zero for $S$-wave states and was omitted
in \cite{FRE02,FRE02b}. Note that in~\cite{ZHOU03} we used a much
simpler spin-orbit interaction which also gives reasonable results
for \ds. The eigenvalue of $M^2_{\rm ho}$, i.e., the mass squared
of a vector meson or a $L\ne 0$ meson, is given by
\begin{equation}
 M^2_{nLJ}
 = 2m_t \left[
         \left( 2n+L+\frac{3}{2}\right) \sqrt{\frac{c_2}{m_r}}
         + \half m_t - c_0 + E_{LJ}
        \right] ,
 n = 0, 1, 2, \cdots .
 \label{eq:masssquare}
\end{equation}
where $E_{LJ}$ gives the spin-orbit splitting.

For pseudoscalar mesons, the Dirac-delta interaction is active
thus the Hamiltonian must be renormalized. One of the
renormalization approaches is the $T$ matrix method~\cite{FRE00}.
The details is found in \cite{FRE02,FRE02b}.

\begin{table}
\begin{center}
\caption{Parameters for the light cone harmonic oscillator model.
$c_0$ and $c_2$ of the harmonic oscillator potential and the
masses of up, down and charm quarks are fixed from the masses of
$\rho(770)$, $\rho(1450)$, $J/\psi(1S)$ and $\psi(2S)$~\protect
\cite{PDG02}, with the assumption of $m_u=m_d$. Strange and bottom
quark masses are determined by the masses of $K^*$ and
$B^*$~\protect \cite{PDG02}. } \label{table:parameters}
\begin{tabular}{lcccccc}
\hline\hline
 Parameters & $c_0$ [MeV] & $c_2$ [GeV$^3$] & $m_u=m_d$ [MeV] & $m_s$ [MeV] &
 $m_c$ [MeV] & $m_b$ [MeV] \\
 \hline
 Values     & $807$       & 0.0713          & 265             & 478         &
 1749        & 5068        \\
\hline\hline
\end{tabular}
\end{center}
\end{table}

The parameters used in the present model are listed in
Table~\ref{table:parameters}. The model was first applied to study
the pseudoscalar-vector splittings in the $S$-wave meson
spectra~\cite{FRE02}. In~\cite{FRE02b} it was used to investigate
the $S$-wave meson spectra from $\pi$-$\rho$ to
$\eta_b$-$\Upsilon$ and to predict top quark meson spectra. The
following conclusions were drawn: (i) the splitting between the
light pseudoscalar and vector meson states is due to the strong
short range attraction in the $^1S_0$ sector; (ii) the linear
relationship between the mass squared of excited states with
radial quantum number~\cite{ANI00} is apparent from our model and
is found to be qualitatively valid even for heavy mesons like
$\Upsilon$; (iii) for the $S$-wave meson spectra from $\pi$-$\rho$
up to $\eta_b$-$\Upsilon$, the simple model presents satisfactory
agreement with available data and/or with the meson mass spectra
given by Godfrey and Isgur~\cite{GOD85}. Therefore, the extension
of the light cone QCD-inspired model which includes a quadratic
confinement while keeping simplicity and renormalizability, gives
a reasonable picture of the spectrum of both light and heavy
mesons.

\section{APPLICATION OF THE LCH MODEL TO THE RESONANCE \ds}
\label{sec:ds}

A recent experiment by the BABAR collaboration observes a new
narrow $c\bar s$ state, \ds\ with the invariant mass $M = 2.317$
GeV/$c^2$~\cite{BABAR03}. Later on this meson is confirmed by the
CLEO Collaboration~\cite{CLEO03}. This state has a natural
spin-parity and the low mass suggests a $J^\pi = 0^+$ assignment.
In the convention of the quark model, correspondingly, $L=1$,
$S=1$ and $J=0$, i.e., $^{2S+1}L_J$ = $^3P_0$. The mass of this
state is typically predicted between 2.4 and 2.6 GeV/$c^2$ in
\cite{GOD85,GK91,DE01} (cf. Table~\ref{tab:comp}). In this
section, we'll apply the light cone harmonic oscillator model to
this and other $P$-wave $D_s$ mesons.

For $L \ne 0$ states, we include phenomenologically a spin-orbit
term as shown in (\ref{mass4}) where the strength for the
spin-orbit interaction $\kappa$ is an additional parameter to be
determined by the data. For $P$ states, the spin-orbit splitting
is derived as
\begin{equation}
 E_{LJ}= \kappa \left( \frac{1}{m_1^2} +  \frac{1}{m_2^2} \right) \times
         \left\{
          \begin{array}{ll}
          \half,     & {\rm for}\ ^3P_2, \\
          \quarter \left(-1+\sqrt{1+2\beta^2}\right), & {\rm for}\ ^1P_1, \\
          \quarter \left(-1-\sqrt{1+2\beta^2}\right), & {\rm for}\ ^3P_1, \\
          -1 ,       & {\rm for}\ ^3P_0, \\
          \end{array}
         \right.
 \label{eq:sosplitting}
\end{equation}
where $ \beta = (m_1^2-m_2^2)/(m_1^2+m_2^2) $. Contrary
to~\cite{ZHOU03}, the spin orbit interaction in (\ref{mass4})
mixes the singlet and triplet states with the same total angular
momentum $J$ when $m_1 \ne m_2$. Therefore the two $J=1$ states,
$^1P_1$ and $^3P_1$ from the pure LS coupling scheme mix with each
other. The larger the difference between the two constituent
masses, the farther $^1P_1$ and $^3P_1$ depart from each other.

\begin{table}[tbp]
\caption{Masses of $P$-wave $c\bar s$ meson states. The data for
$^3P_2$ and $^1P_1$ are taken from~\protect\cite{PDG02} and that
for $^3P_0$ from~\protect\cite{BABAR03,CLEO03}. Predictions
from~\protect\cite{GK91,DE01} are also included for comparison.
For the labels of each state, $n$ is the radial quantum number,
$L$ the total orbital angular momentum, $S$ the total spin and $J$
the total angular momentum. The second row gives the notation
given in~\protect\cite{DE01} where $j$ is the total angular
momentum of the strange quark which is conserved in the limit of
large charm quark mass. In the present work, the mass of $^3P_2$
(included in a square braket) is used to fix the parameter for the
spin-orbit coupling.} \label{tab:comp}
\begin{center}
\begin{tabular}{cccccc}
\hline\hline
 $n^{2S+1}L_J$ & $n^jL_J$   & Data  & Ref.~\protect\cite{GK91} &
 Ref.~\protect\cite{DE01} & This work  \\
\hline
 $0^3P_2$ & $0^{3/2}P_2$ & 2.573 & 2.59 & 2.581 & [2.573] \\
 $0^1P_1$ & $0^{3/2}P_1$ & 2.536 & 2.55 & 2.535 & 2.517 \\
 $0^3P_1$ & $0^{1/2}P_1$ & ---   & 2.55 & 2.605 & 2.387 \\
 $0^3P_0$ & $0^{1/2}P_0$ & 2.317 & 2.48 & 2.487 & 2.327 \\
\hline\hline
\end{tabular}
\end{center}
\end{table}

The mass of $D_s(^3P_2)$, 2.573 GeV~\cite{PDG02} is used to
determine the parameter $\kappa = 0.03842$ GeV$^3$. The masses of
the other three $P$ states are easily calculated from
Eqs.~(\ref{eq:masssquare}) and (\ref{eq:sosplitting}) and given in
Table~\ref{tab:comp} where predictions from~\cite{GK91,DE01} are
also included for comparison.

The LCH model reproduces the available data very well. The data
are available for other two $P$-wave $c\bar s$ states except
$D_s(^3P_2)$ the mass of which is used to determine the parameter
$\kappa$ for the spin-orbit interaction. The calculated $^1P_1$
mass deviates from the datum by only 0.019 GeV. Remarkably, the
present prediction for the mass of the lowest $P$ state \ds\ is in
good agreement with the experimental value. Therefore from our
model, \ds\ might still be a ``normal'' meson consisting of two
constituent quarks which agrees with our previous
conclusion~\cite{ZHOU03} and many other recent works~\cite{BEV03}.

\section{SUMMARY}
\label{sec:summary}

We applied the light cone harmonic oscillator (LCH) model---a
renormalized light cone QCD-inspired effective theory with a
quadratic confinement in the mass squared operator---to study the
meson spectra.

The model was applied to the $S$-wave mesons from $\pi$-$\rho$ up
to $\eta_b$-$\Upsilon$~\cite{FRE02,FRE02b}. In this model, the
splitting between the light pseudoscalar and vector meson states
is due to the strong short range attraction in the $^1S_0$ sector.
The linear relationship between the mass squared of excited states
with radial quantum number~\cite{ANI00} is naturally explained.
This model presents reasonable agreement with available data.

The $P$ states of the charmed strange meson are investigated by
using the LCH model with a phenomenological spin-orbit interaction
included (cf.~\cite{ZHOU03}). The mass of the recently found
meson, \ds~\cite{BABAR03,CLEO03} could be reproduced quite well by
this simple model.

\section*{ACKNOWLEDGEMENT}

I thank Professors Hans-Christian Pauli and Tobias Frederico for
helpful discussions and fruitful collaborations. I am indebted to
Professor Shun-Jin Wang for valuable discussions on the spin-orbit
interaction. I would like also to thank Dr. Simon Dalley and the
other organizers of this meeting for their warm hospitality in
Durham. This work was partly supported by the Major State Basic
Research Development Program of China Under Contract Number
G2000077407 and the National Natural Science Foundation of China
under Grant Nos. 10025522, 10221003 and 10047001.

\end{document}